# An Infrastructure Cost Optimised Algorithm for Partitioning of Microservices

KALYANI V N S PENDYALA

The Cloud Computing and Distributed Systems (CLOUDS) Laboratory, School of Computing and Information Systems, The University of Melbourne, Australia

kpenyala@student.unimelb.edu.au

RAJKUMAR BUYYA

The Cloud Computing and Distributed Systems (CLOUDS) Laboratory, School of Computing and Information Systems, The University of Melbourne, Australia

rbuyya@unimelb.edu.au

The evolution and advances made in the field of Cloud engineering influence the constant changes in software application development cycle and practices. Software architecture has evolved along with other domains and capabilities of software engineering. As migrating applications into the cloud is universally adopted by the software industry, microservices have proven to be the most suitable and widely accepted architecture pattern for applications deployed on distributed cloud. Their efficacy is enabled by both technical benefits like reliability, fault isolation, scalability and productivity benefits like ease of asset maintenance and clear ownership boundaries which in turn lead to fewer interdependencies and shorter development cycles thereby resulting in faster time to market. Though microservices have been established as an architecture pattern over the last decade, many organizations fail to optimize the architecture design to maximize efficiency. In some cases, the complexity of migrating an existing application into the microservices architecture becomes overwhelmingly complex and expensive. Additionally, automation and tool support for this problem are still at an early stage as there isn't a single well-acknowledged pattern or tool which could support the decomposition. This paper discusses a few impactful previous research and survey efforts to identify the lack of infrastructure cost optimization as a parameter in any of the approaches present. This paper proposes an Infrastructure-optimised predictive algorithm for partitioning monolithic software into microservices. It also summarizes the scope for future research opportunities within the area of microservices architecture and distributed cloud networks.

**CCS CONCEPTS** • [Information systems applications](#) • [Enterprise information systems](#) • [Enterprise resource planning](#)

**Keywords—**partitioning, infrastructure optimization, software architecture, cloud applications, microservices.

## I. INTRODUCTION

The famous quote "Change is the only constant" is most applicable in the field of software engineering and technology. The drift and shift in the technology landscape is continuous due to the rich contributions towards advancements in the fields of distributed computing, cloud deployment offerings, software development frameworks and more. These advancements have led to a quest for more scalable, resilient, reliable, and efficient software applications. To support these innovations, the software industry is continuously developing techniques and processes to improve the application performance along with the performance of the software development teams. As the agility of applications is growing the need to adopt new architectural designs is gaining prominence as it is the backbone of the performance characteristics [2] of a software-intensive system.

The software architecture of a system depicts the organisation or structure of the system and explains how it behaves. An efficient architecture is the foundation for efficient software. [3] As software architecture gives a visual depiction of the system to be built with all its components and behaviours, it allows the architect or designer to perceive possible outcomes and shortcomings. Consequently, it ensures the final built product is reliable both from its functional outcomes and other efficiency parameters considered.

Given the importance of software architecture, earlier software
development models treated the architecture, implementation and operations of software to be completely independent components. however, within in last decade, there has been a huge shift in the traditional operational style. Software architecture has broadened its definition to a combination of elements, behaviours, and design concepts and it must consider the extensibility and flexibility of the software as the change in scope of software is very less predictable.

This work encapsulates the history of software architecture evolution, advantages and challenges with Microservices architectural style and focuses on optimisation of the microservices decomposition strategy. As identified in the extensive survey [1] the decomposition guidelines and tools are in the early stage as there aren't any widely accepted or successful standards defined. So, this work proposes a monolith to microservices partitioning approach where a monolith software is represented collectively as a set of existing source code, functional flows, and infrastructure resource templates and the partitioning of the monolith into microservices targets the optimisation of infrastructure cost and operational efficiency. The proposed method is evaluated and compared with a few of the existing approaches. The paper concludes by identifying the open research opportunities in the microservices architecture domain.

## II. SOFTWARE ARCHITECTURE EVOLUTION

Building software is analogous to building a house. the building of any house starts with planning. It involves an architect providing a blueprint or architecture document to help visualise the final built product and ensure all the functional capabilities and connectivity requirements are captured. software architecture [4] plays a similar role in the software development life cycle, with emerging trends of technology and cloud offerings, software architecture patterns have also evolved as shown in Figure 1.

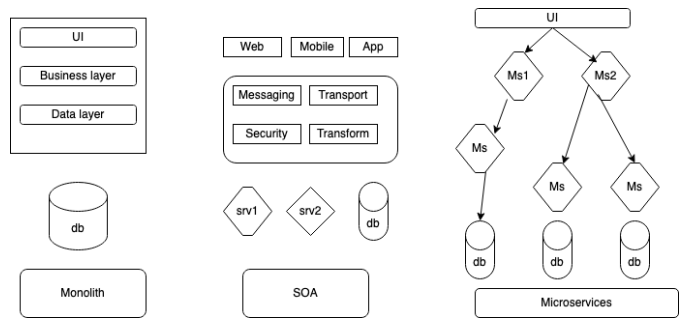

Figure - 1 Software Architecture Evolution.

One early pattern is a Monolithic software architecture which is a traditional unified container of whole application functionality and often comprises of a single large deployable unit packaged with all its dependencies and originating from a single complex code base being shared between multiple development teams. As the application grows, the monolith architecture faces several challenges as it is difficult to scale, and hard to maintain and enhance. This further leads to operational complexity, absence of fault isolation mechanism and hard software coupling, resulting in long incident management-related downtimes and long build/test/release cycles.

The next major shift in the architectural styles is with layered software architecture where the application is majorly layered based on the responsibility of the module where each layer has dedicated responsibilities like Client-Server architecture [5]. A client-side user interface, a server-side program, and/or a database are often included in such a system. Followed by SOA (service-oriented Architecture) [6] where each service provides a business capability that reduces the duplication of functionality and encourages reusable components. SOA architecture improves the time to market due to the modularisation capabilities provided.

Microservices are an evolution of SOA architecture consisting of a collection of small, autonomous services [7], Each service is self-contained, self-deployable and ideally should implement a single business capability within a bounded context. A bounded context is a natural division within a business and provides an explicit boundary within which a domain model exists.

A few advantages the microservices architecture brings in are Ease to Build, Ease to Enhance, Ease to deploy, selective scalability where only selected services with higher traffic could be scaled accordingly, Fault Isolation which enables partial run states of the application, smaller release cycles, Dependencies are specific and optimal, increased application and data security. All these attractive benefits made Microservices an anonymous choice of architecture for many organizations especially those that chose to be cloud native or migrating to cloud.

While building a microservices application from scratch could be clean and enable the organizations with the ideal benefits of the architecture, the most practical approach would be to migrate existing legacy monolithic applications into microservices architecture and this problem of determining the boundaries of each microservice extracted from the monolith and defining the functional scope of each microservice is known as microservices partitioning or decomposition.

The partitioning task plays an important role in determining the efficiency and performance of the implemented application and also plays a key role in enabling the application to fetch all the benefits that are provided by the architecture pattern, though the definition of the partitioning task is straight forward - the partitioning task needs to divide a whole set of functional capabilities and a unified source code base into individual services which as a whole should satisfy all the functional flows delivered by the monolith. In simple terms, the problem can be looked at as dividing a set of classes into multiple small subsets satisfying some conditions. The solutions and automation tools available for this problem are still in their infancy.

## III. RELATED WORK

An efficient innovation or research contribution would be only possible when the current state of the art and problem domain are well understood, and the research gaps are evaluated.

Microservice identification and extraction is an open challenge in the software industry which is currently being solved using human intellect and domain excellence. There were a few approaches proposed earlier to automate the whole process of microservice extraction. Where many papers have elaborated on the advantages of microservice architecture, there are also many challenges in implementing [8, 9] them optimally even when done manually following many evolving guidelines.

As mentioned, [9] many of the challenges are caused by the increased number of moving parts in the system. This architectural style also introduces organizational operational challenges like splitting teams and designating responsibilities. As part of their work, EMIL & ERIK have identified a few important factors to be considered in the partition process identifying Service boundaries, Automation in integration and deployment, Communication between services, Decentralized data, Fault tolerance and fault handling.

These factors need to be considered as prime metrics when any tool, is proposed or implemented for the partitioning service. To efficiently develop an algorithm and implement we need a standard representation of the existing system and the desired result set of systems. We also need to define the metrics to be optimized in this process, input parameters, and standard output characteristics to allow performance comparison to existing approaches.

An extensive literature review and methodological survey have been conducted [1] which defines the microservices partitioning work has been broadly phased into Input collection, Monolith Analysis, Microservices identification, Microservices optimisation, Evaluation, and Deployment shown in Figure 2. Though few tools have targeted a few phases independently there is no unique standard tool that is widely evaluated and accepted.

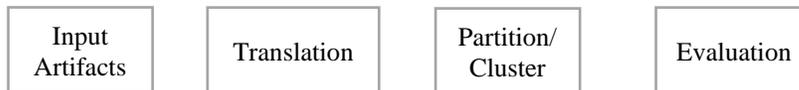

Figure - 2 Phases of partitioning task.

The input collection is an important step as it represents the existing monolith application, acts as the current knowledge set and is a key for visualisation and efficient modelling. The inputs used across the significant works are code base, [10] [11] [12]domain models, application real-time and performance logs [13], functional flows [14]identified by test cases, and application traces [15]. These inputs of an application could be collected using static and dynamic analysis tools [16], Once the input is collected representing the application or modelling this collected input data would be the next step, The most suitable representation adapted for a software application is a graph [11] and some approaches use SET representation.

The next step would be Microservices identification or partitioning FoSCI [15] uses the monolith's execution traces as the input, and extracts service candidates using a search-based functional atom grouping algorithm. Bunch [12] relies on source code to extract the graph of the entities (e.g., classes) and relations (e.g., function calls) in the source code and applies search-based clustering algorithms to produce a subsystem.

Mono2Micro [14] performs spatiotemporal decomposition, leveraging well-defined business use cases and runtime call relations to create functionally cohesive partitioning of in application classes. This approach uses hierarchical clustering using jacquard distance, predefining the number of subsystems or clusters that are desired from the partitioning.

CO-GCN [11]is the very early approach where deep learning is used for microservice clustering, CO-GCN uses the static code-derived Graph as a feed to a graph convolutional neural network and supervised learning techniques are applied to populate subgraphs which could be the potential microservice candidates.

DEEPLY [17] is an extension on the CO-GCN where a similar Graph-based CNN approach is proposed whereas the focus of this work was mainly on fine-tuning the hyper hyper-parameter optimizers (HPO) to improve the performance of the predefined metrics.

Evaluating the extracting microservices is also challenging as there are no concrete metrics or evaluation benchmarks available, though there are few open source monolith applications whose source code is available and are used as benchmark candidates across past research work, comparison of each of the methods to the other is not accurate as the evaluation parameters used and the definition of the evaluation parameters vary from implementation to implementation, generic evaluation metrics used are Modularity, Interface Number, Cluster sizing, Inter-service communication, [11], coupling and cohesion are also popular metrics [18]. Table 1 notes a summary of the focus of past works.

**Table 1**: Overview of techniques in literature

| Input | Partitioning techniques | Benchmark applications | Evaluation parameters |
|---|---|---|---|
| Source code | Clustering (hierarchical, kmeans) | DayTrader | cohesion |
| Build artifact | Genetic algorithms | PBW | coupling |
| Functional logs, execution traces | Neural networks | AcmeAiR | Inter service calls |
| Performance logs | Dependency graph analysis | Dietapp | Cluster size |
| Domain model | Cogcn, deep learning | JpetStore | |
| Dataflow diagrams | Graph partitioning algorithms | | |
| Test cases /scenarios | | | |

## IV. PROPOSED ALGORITHM

From an extensive review of the current literature and exploring the industry pain points, we identified that many organizations struggle to efficiently partition and migrate an existing monolith into microservices as there are not enough automated tool support or visualisation/ modelling tools available which help architects simulate the microservices applications. As we migrate from one single deployable application into multiple unified deployable units, Total infrastructure cost is one of the important parameters that needs to be considered and optimised when designing the partitions which has not been considered in any of the past works.

In this paper, we propose using total infrastructure cost as a significant parameter in the partitioning algorithm. In general, microservices partitioning realisation has three major steps:

**Step 1**: Choosing the input artifacts and translating the input artifacts into analytical data format, as part of this work we chose three different artifacts related to an application source Code which is the base input that needs to be partitioned into sub-sets of classes, provides all the dependency information and relationships between different classes, the stack trace of the application for one week to make sure all the business cases are covered, Current infrastructure file. We have used a benchmark open-source Java monolith daytrader which has a total of 111 Java classes.

Once the input artifacts are all gathered, the Translation of this input into the form of graphs is done using a Class dependency analyzer on the source code, we have based the graph translation on [11] a custom rule-based engine we defined to translate the stack trace into functional dependency matrix. As the scope of this work is to optimise the infrastructure cost, the parameters affecting this cost must be properly defined, we have used the YAML file of the existing open-source application and implemented a few rules to abstract the template of cloud resources that existing application is using and define the parameter of infrastructure as the total number of infra components that the monolith application is using and we also define the future state infrastructure cost by tagging the class files using each of the components and thus be able to predict the total infrastructure cost of the split microservices.

The current resource factor is defined as:

Infrastructure Factor = (Nvms, Nfs, Ndb, Nca)

While these parameters respectively Number of deployable VMs, Number of file storages, Number of database nodes, and Number of cache instances are mapped to the Number of Ec2, Number of S3, Number of databases, Number of Caches required in future dates based on the Algorithm 1.

ALGORITHM 1: Predictive Infrastructure

**Algorithm 1: Predictive infrastructure factor definition**
**Input: Class Nodes (P), resource Nodes (R), yml**
**Output: IF = (Nec, Ns3, Ndb ,Nca)**
**1 for each** resource r in R
2    Function(P,R)
3.       If (exists a resource edge from Pi)
4.           identify the infrastructure resource type from yml
5.           translate the resource to the cloud infrastructure component
6.       (if the type of R is file storage, Ns3 = Ns3+1)
6.    consolidate the total number of predictive resources
7.    Output the Infrastructure factor.

Each class in the application source code is denoted by a Node P and the class dependency analyser is used to define the Edges between the P nodes, an edge is present with Pij to Pnj if there is a direct dependency from class I to class N, all the resources of the application ( file storage, database, ec2 instances, caches) are represented as Resource nodes and an edge between a resource node Rij to Pin exist if the class N depends on the resource I, then we process the execution traces sorted and grouped into functional flows so each set of execution traces denote one complete functional flow, so we build a FF graph where all the classes part of a functional flow Fi would be having an entry/ edge FiPi.

So as a whole the application is represented by G= (P,R,F,Er,Ef) a sample representation can be seen in Figure 3.

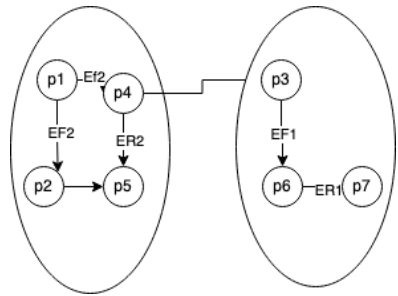

Figure - 3 Sample Application Representation.

**Step 2**: Choosing the right partitioning algorithm with partitioning rules is the second step in the process, after exploring the available graph partitioning and clustering algorithms, we have chosen an algorithm as in general the graph partitioning algorithms try to keep the interactions between the partitions minimal and that is in line with one of our partitioning objectives i.e. to keep the number of inter-partition calls (IPC) at a minimum, As the graph partitioning problem is an NP-Hard and there is no single perfect algorithm/ solution we have used METIS [19], a weighted graph partitioning algorithm to partition the Weighted graph created in step 1 with 'TOI$\alpha$' to be optimised.

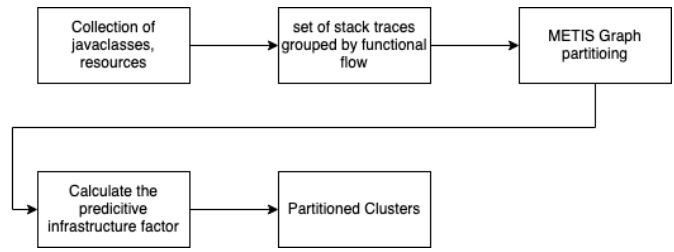

Figure - 4 Proposed implementation steps.

**Step 3:** Now that we have the partitions and microservices split from the monolith evaluating the performance or the efficiency of partitions created in step 3, though there is no single way to evaluate our implementation of the existing approaches as none of the previous works has considered infrastructure optimisation as a partitioning objective, we still used the Newman Girvan Modularity [20] 'NGM' metric to measure the quality of the partitions created and the number of microservices generated.

Our implementation steps, as represented in Figure 4, have provided an NGM score of 0.5, with the number of microservices as 4 for the day trader monolith application.

## V. IMPLEMENTAION AND EVALUATION

As discussed, the implementation of the proposed approach can be divided into Data collection, Data Translation, Partitioning, Evaluation This section describes each of these steps in detail.

**Dataset**

We have chosen four popular benchmark open-source applications namely daytrader, jpetstore, springBlog and PBW. whose properties are summarized in Table 2. All these monolith applications are Java applications. The source code of the applications and deployment configuration files are the data sources for the implementation. The application.yml file available for the SpringBlog application makes it suitable for the infrastructure property configuration.

**Table 2**: Dataset properties

| Application | classes | clusters | Infrastructure Involved |
|---|---|---|---|
| Daytrader https://github.com/WASdev/sample.daytrader7 | 111 | 6 | Compute, database |
| Jpetstore https://github.com/mybatis/jpetstore-6 | 24 | 3 | Compute, database |
| springBlog https://github.com/Raysmond/SpringBlog | 47 | 5 | Compute, database, cache |
| PBW https://github.com/WASdev/sample.daytrader7 | 36 | 4 | Compute, database, |

**Data translation**

Once the data sources are collected conversion of a Java application into a graph representation defining the various nodes, class nodes, and resource nodes, building the dependency graph and defining the weights of the edges based on the infrastructure configuration is part of the data translation step.

We used the class dependency analyzer (CDA) [21] tool to extract the dependencies and relationships between the classes, when CDA is run on an application jar file an XML file with the class-to-class relationships is generated.

We have implemented a rule filter to convert the XML generated by the CDA into a graph where the nodes of the graph represent the Application classes and Edges defining the relationship between the nodes, the weight of an edge is defined based on the nature of the relation and depending on the common infrastructure components between the nodes.

If N1 and N2 share an infrastructure resource DB1 the weight parameter of edge E12 is incremented by 1. Thus, an application graph is defined in total.

**Data Partitioning**

The graph partitioning algorithm used is METIS [19], METIS is proven to generate partitions with the minimal number of cross-partition edges and is an efficient algorithm for partitioning weighted graphs.

Figure 5 shows the result of the partitioning of the jpetstore application.

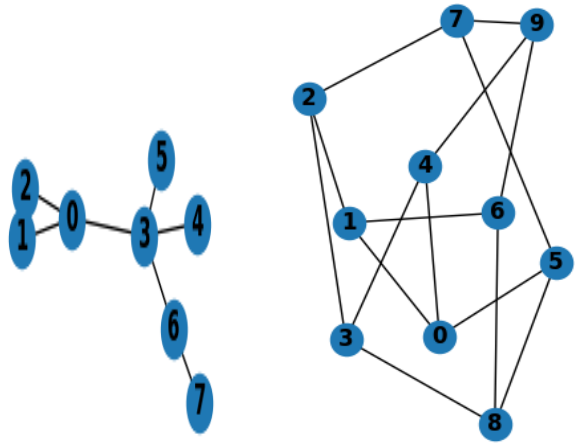

Figure - 5 Resulting partitions of the Jpetstore application.

**Performance Evaluation**
To evaluate our proposed solution, we have chosen three parameters that are recommended in the literature to measure the effectiveness of the partitions defined by our approach. The parameters are F-measure, NGM Modularity, and Interface Number (IFN).

**F-measure**

F-measure or F-score is used to evaluate the accuracy and quality of the clusters defined, the accuracy metric computes how many times a model made a correct prediction across the entire dataset it is a harmonic mean usually of precision and recall and ranges between 0 and 1. F1 measure is calculated as

$$F1 = (precision * recall)/(precision + recall)$$

**Newman Girvan Modularity**

The Girvan-Newman algorithm is popularly used to evaluate the modularity of a graph, the higher the NGM value higher the significance of the community structure. This measure identifies the density of clusters based on the principle that random nodes in a graph wouldn't be densely connected as a high modular graph or cluster. It compares the same edges of a cluster to the partition members of randomly generated graph edges.

**Interface Number**

Interface Number is a direct indication of the interactions between the clusters or classes, minimum IFN number between the clusters would mean the clusters or microservices obtained are optimal and need less inter-communication between the microservices.

The evaluation results on the datasets chosen are listed in Table 3. F1 score measure is very specific to the context, but a general rule is an F1 measure of 0.7 or higher is good, as we can see the values on Jpetstore, and spring blog applications are in the good range which indicates the algorithms can generate partitions with balance. Similarly, the NGM ranges are from -0.5 to 1 higher the value of the NGM score significant is in the community structure, in our experiment though the Springblog application has NGM on the lower side PBW applications have a score close to 1.

**Table 3**: Evaluation Results on Datasets

| Dataset | F1 | NGM |
|---|---|---|
| JPetstore | 0.7 | 0.44 |
| daytrader | 0.6 | 0.5 |
| springblog | 0.8 | 0.34 |
| pbw | 05 | 0.9 |

## VI. CONCLUSIONS AND FUTURE SCOPE

This research work discussed the need for microservices partitioning algorithms and their current state of the art. It also proposed the need for considering the infrastructure cost as an impact parameter to decide the microservices partitioning. The method tests the proposed approach with a benchmark monolith application and identifies future extensions possible to this work. as any automation and simulation of microservices partitioning would be a huge saving of initial investment to the organizations, supporting them in designing their microservices architecture effectively. all the benefits of the popular architecture pattern could be availed with an eye on cost/resource optimization.

As part of the research work, we identified the amount of tool and automation support available for data representation and translation is less. there are no tools easily available for converting an existing monolith application into a graph format or tools to analyze the stack trace, especially if the logging tools used are legacy as is the general case with any monolith applications being subject to partition.

Currently, there is rare to no emphasis on optimization of resources and performance metrics to be considered as impacting parameters for microservices and partitioning approaches. Therefore, investigating predictive artificial intelligence algorithms and data analysis to the partitioning approaches will help in predicting approximate cost and allow cost-efficient microservices.

Another research area open for exploration is the evaluation parameters definition and benchmarking as the evaluation parameters for the microservices partitioning task are still vague in definition and there are no unique open-source benchmark data available for the researchers to validate and evaluate their work against. This would save the initial investment and aid the organizations design their microservices architecture effect.

**BIBLIOGRAPHY**

[1] Yalemisew, Abgaz and McCarren, Andrew , "Decomposition of Monolith Applications Into Microservices Architectures: A Systematic Review," *IEEE Transactions on Software Engineering,* vol. 49, 2023.

[2] M. Shaw and P. Clements, "The golden age of software architecture," *IEEE Software,* 2006.

[3] IEEE Standards, "IEEE Recommended Practice for Architectural Description for Software-Intensive Systems," *IEEE Std 1471-2000,* 2000.

[4] David Garlan and Mary Shaw, "An introduction to software architecture.," *Advances in Software Engineering and Knowledge Engineering.*

[5] Kassab, Mohamad & Mazzara, Manuel & Lee, JooYoung & Succi, Giancarlo. , "Software architectural patterns in practice: an empirical study," *Innovations in Systems and Software Engineering,* 2018.

[6] Erickson, John & Siau, Keng. , "Service Oriented Architecture: A Research Review from the Software and Applications Perspective," *Innovations in Information Systems Modeling: Methods and Best Practices,* 2009.

[7] H. Zhang, S. Li, Z. Jia, C. Zhong and C. Zhang,, Microservice Architecture in Reality: An Industrial Inquiry, 2019 IEEE International Conference on Software Architecture (ICSA), 2019.

[8] Tozzi, Christopher, "6 Reasons Not to Adopt Microservices," [Online]. Available: https://cloudnativenow.com/features/microservices-use-not-use-question/.

[9] Benjamin Benni, Sébastien Mosser, Jean-Philippe Caissy, "Can microservice-based online-retailers be used as an SPL? a study of six reference architectures.," in *ACM Conference on Systems and Software Product Line*.

[10] Vikram Nitin, Shubhi Asthana, Baishakhi Ray, and Rahul Krishna., "CARGO: AI-Guided Dependency Analysis for Migrating Monolithic Applications to Microservices Architecture," *IEEE/ACM International Conference on Automated Software Engineering,* 2023.

[11] Desai, Utkarsh & Bandyopadhyay, Sambaran & Tamilselvam, Srikanth., "Graph Neural Network to Dilute Outliers for Refactoring Monolith Application," *AAAI Conference on Artificial Intelligence,* 2021.

[12] S. Mancoridis, B. S. Mitchell, Y. Chen and E. R. Gansner, "Bunch: a clustering tool for the recovery and maintenance of software system structures," in IEEE *International Conference on Software Maintenance* , 1999.

[13] Y. Zhang, B. Liu, L. Dai, K. Chen and X. Cao,, "Automated Microservice Identification in Legacy Systems with Functional and Non-Functional Metrics," in IEEE *International Conference on Software Architecture (ICSA)*, 2020.

[14] Anup K. Kalia, Jin Xiao, Rahul Krishna, Saurabh Sinha, Maja Vukovic, and Debasish Banerjee, "Mono2Micro: a practical and effective tool for decomposing monolithic Java applications to microservices," in *ACM Joint Meeting on European Software Engineering Conference and Symposium on the Foundations of Software Engineering*, 2021.

[15] W. Jin, T. Liu, Y. Cai, R. Kazman, R. Mo and Q. Zheng, "Service Candidate Identification from Monolithic Systems Based on Execution Traces," *IEEE Transactions on Software Engineering,* 2021.

[16] Matias, T., Correia, F.F., Fritzsch, J., Bogner, J., Ferreira, H.S., Restivo, A., "Determining Microservice Boundaries: A Case Study Using Static and Dynamic Software Analysis," *springer,* 2020.

[17] Rahul Yedida, Rahul Krishna, Anup Kalia, Tim Menzies, Jin Xiao, and Maja Vukovic, "An expert system for redesigning software for cloud applications," in *Expert Syst. Appl.*, 2023.

[18] Shanshan Li, He Zhang, Zijia Jia, Zheng Li, Cheng Zhang, Jiaqi Li, Qiuya Gao, Jidong Ge, Zhihao Shan,, "A dataflow-driven approach to identifying microservices from monolithic applications," in *Journal of Systems and Software,*, 2019.

[19] Karypis, George & Kumar, Vipin., "METIS: A software package for partitioning unstructured graphs, partitioning meshes, and computing fill-reducing orderings of sparse matrices," 1997.

[20] Newman, Mark & Girvan, Michelle., "Finding and evaluating community structure in networks," in *E, Statistical, nonlinear, and soft matter physics*, 2004.

[21] Manfred Duchrow, "Class Dependency Analyzer," [Online]. Available: http://www.dependency-analyzer.org/.

[22] M. S. Tamboli, "daytrader," [Online]. Available: https://github.com/WASdev/sample.daytrader7.

[23] L. Dobrica and E. Niemela, "A survey on software architecture analysis methods," *IEEE Transactions on Software Engineering,* 2002.